\documentclass{article}

\usepackage{arxiv}

\usepackage[utf8]{inputenc} 
\usepackage[T1]{fontenc}    
\usepackage{hyperref}       
\usepackage{url}            
\usepackage{booktabs}       
\usepackage{amsfonts}       
\usepackage{nicefrac}       
\usepackage{microtype}      
\usepackage{lipsum}
\usepackage{graphicx}
\usepackage{longtable}
\graphicspath{ {./images/} }

\title{SarcGraph for High-Throughput Regional Analysis of Sarcomere Organization and Contractile Function in 2D Cardiac Muscle Bundles}

\author{
 Saeed Mohammadzadeh \\
  Department of Mechanical Engineering\\
  Boston University\\
  Boston, Massachusetts \\
  \texttt{saeedmhz@bu.edu} \\
   \And
 Yao-Chang Tsan \\
  Department of Internal Medicine\\
  University of Michigan\\
  Ann Arbor, Michigan \\
  \texttt{yctsan@umich.edu} \\
  \And
 Aaron Renberg \\
  Department of Internal Medicine\\
  University of Michigan\\
  Ann Arbor, Michigan \\
  \texttt{renberg@med.umich.edu} \\
  \And
 Hiba Kobeissi \\
  Department of Mechanical Engineering\\
  Boston University\\
  Boston, Massachusetts \\
  \texttt{hibakob@bu.edu} \\
  \And
 Adam Helms \\
  Department of Internal Medicine\\
  University of Michigan\\
  Ann Arbor, Michigan \\
  \texttt{adamhelm@med.umich.edu} \\
  \And
 Emma Lejeune \\
  Department of Mechanical Engineering\\
  Boston University\\
  Boston, Massachusetts \\
  \texttt{elejeune@bu.edu}
}

\begin{document}

\maketitle

\begin{abstract}
Timelapse images of human induced pluripotent stem cell-derived cardiomyocytes (hiPSC-CMs) provide rich information on cell structure and contractile function. However, it is challenging to reproducibly generate tissue samples and conduct scalable experiments with these cells. The two-dimensional cardiac muscle bundle (2DMB) platform helps address these limitations by standardizing tissue geometry, resulting in physiologic, uniaxial contractions of discrete tissues on an elastomeric substrate with stiffness similar to the heart. 2DMBs are highly conducive to sarcomere imaging using fluorescent reporters, but,  to their larger and more physiologic sarcomere displacements and velocities, prior sarcomere-tracking pipelines have been unreliable. Here, we present adaptations to SarcGraph, an open-source Python package for sarcomere detection and tracking, that enable automated analysis of high-frame-rate 2DMB recordings. Key modifications to the pipeline include: 1) switching to a frame-by-frame sarcomere detection approach and automating tissue segmentation with spatial partitioning, 2) performing Gaussian Process Regression for signal denoising, and 3) incorporating an automatic contractile phase detection pipeline. These enhancements enable the extraction of structural organization and functional contractility metrics for both the whole 2DMB tissue and distinct tissue regions, both in a fully automated manner. We complement this software release with a dataset of 130 example movies of baseline and drug-treated samples disseminated through the Harvard Dataverse. By providing open-source tools and datasets, we aim to enable high-throughput analysis of engineered cardiac tissues and advance collective progress within the hiPSC-CM research community.
\end{abstract}

\keywords{SarcGraph \and hiPSC-Cardiomyocytes \and Sarcomere \and Contractility \and Image Analysis \and High-Throughput Analysis \and 2D Cardiac Muscle Bundle \and Open-source Software}

\section{Introduction}
Human induced pluripotent stem cell-derived cardiomyocytes (hiPSC-CMs) have enabled breakthroughs in cardiovascular research (Mohr et al., 2022). This technology provides a valuable tool for investigating human cardiac development (van den Berg et al., 2015), disease mechanisms (Burridge et al., 2016), and drug testing (Gintant et al., 2019). In particular, when assembled into spontaneously beating tissues in vitro, hiPSC-CMs enable advanced and subject-specific disease modeling (Sacchetto et al., 2020), exemplified by work in high-throughput drug screening (Honda et al., 2021), and investigations of developmental and regenerative processes (Huang et al., 2020). 

However, the structural and functional immaturity of hiPSC-CMs, particularly their relatively disorganized mechanics compared with cardiomyocytes in the adult myocardium (Bedada et al., 2016), can make analyzing both the structural and functional behavior of hiPSC-CMs quite challenging. For example, existing analysis methods struggle with the circular morphology and nonlinear contraction patterns of immature hiPSC-CMs, making it challenging to accurately track sarcomere function (Toepfer et al., 2019) or quantify structural organization in cells that exhibit disorganized structures (Gerbin et al., 2021). In addition, variability in both tissue fabrication platforms and downstream measurement protocols can make hiPSC-CM experiments difficult to reproduce, thus hindering collective progress (Pioner et al., 2016). In order to make hiPSC-CM in vitro tissue experiments both reproducible and suitable for high-throughput experiments, it is necessary to create new hardware and software tools specifically focused on enabling automatic large-scale assessment (Ewoldt et al., 2025). 

To address these challenges, Tsan et al. (2021) previously developed a two-dimensional cardiac muscle bundle (2DMB) platform that standardizes tissue geometry while retaining high-throughput compatibility. In contrast with traditional single-cell monolayers, the 2DMB platform enables generation of purified cardiomyocyte tissues that are highly aligned and able to develop physiologic levels of fractional shortening since 1) each tissue contracts on an underlying elastic substrate with stiffness approximating the heart (~8 kPa), and 2) the geometry of each tissue (7:1 aspect ratio rectangles) induces robust myofibillar alignment, resulting in a uniaxial contractile direction. Additionally, each 2DMB is mechanically decoupled from neighboring tissues (each tissue is ~308 $\mu m$ length and ~44 $\mu m$ width with a buffering space between tissues of 120 $\mu m$ in the long-axis direction and 80 $\mu m$ in the short axis direction). In contrast to 3D engineered tissues, 2DMBs are composed of purified cardiomyocytes, enabling precise study of cardiomyocyte-intrinsic function, and are vertically thin (~10 $\mu m$) making them highly conducive to live-cell sarcomere imaging. The platform thus enables high-throughput, reproducible analysis of contractile function, effects of pharmacologic perturbations, and effects of genetic variants that alter sarcomere function. In the original work, quantification of contraction dynamics was performed by feature tracking from brightfield microscopy to estimate whole-tissue shortening (Tsan et al., 2021). Myofibrillar density was also quantified using live cell actin labeling, but the resolution was not sufficient to reliably detect and measure individual sarcomeres. Here, we have have integrated an $\alpha$-actinin-GFP reporter hiPSC-CMs to clearly resolve individual sarcomeres in contracting cardiac muscle bundles (Figure 1A), enabling more granular analysis that captures previously inaccessible information including spatial variations in contractile behavior, asynchronous activation, and variations in sarcomere organization.

\begin{figure}[ht]
    \centering
    \includegraphics[width=0.95\textwidth]{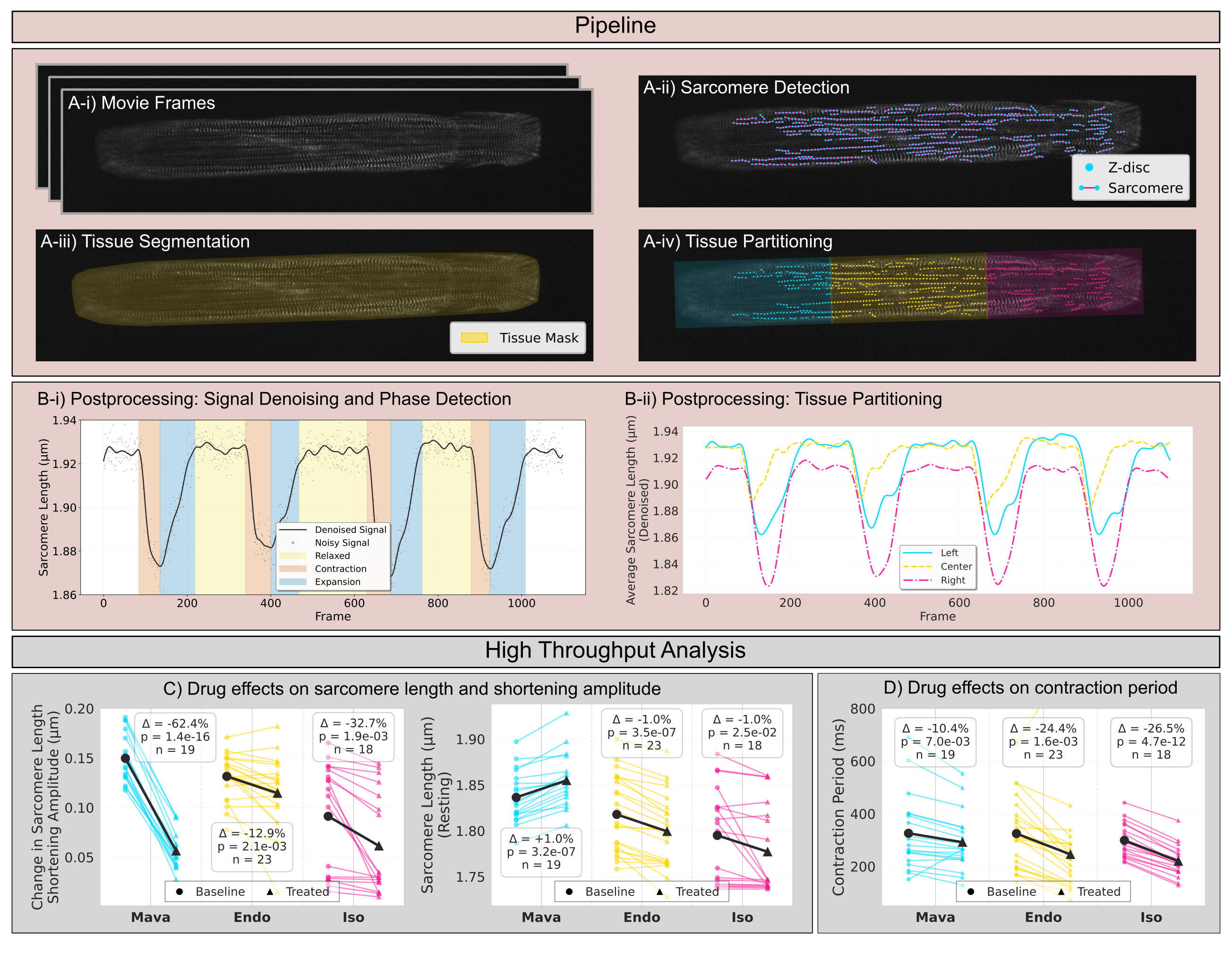}
    
    \caption{\textbf{An overview of modified “SarcGraph” pipeline for high-throughput structural and contractile analysis of 2D cardiac muscle bundles:} (a) The image analysis pipeline is designed to (i) processes high-frame-rate videos of beating 2D cardiac muscle bundles , (ii) conduct sarcomere detection, (iii) tissue segmentation, and (iv) tissue partitioning. (b) The second step of the pipeline processes the sarcomere length signals for contractile phase detection and regional analysis, in this figure we (i) show an example of Gaussian process regression for signal denoising, with detected contractile phases (relaxed in yellow, contraction in orange, expansion in blue) overlaid on the sarcomere length signals, and (ii) show an example of regional sarcomere length signals from three longitudinal tissue segments, demonstrating spatial heterogeneity in contractile behavior across the tissue. This pipeline enables high-throughput analysis, for example: (c) paired line plots displaying changes in sarcomere shortening amplitude (left) and resting sarcomere length (right) for individual samples as well as the mean for each group under three drug treatments (mavacamten, endothelin-1, and isoproterenol), with percentage changes ($\Delta$ = percent change in mean from baseline to treated), p-values from paired t-tests indicating statistical significance of the drug effect, and sample sizes (n = number of paired samples) noted, and (d) paired line plots displaying contraction period changes for individual samples as well as the mean for the group under the same three drug treatments, with percentage changes ($\Delta$ = percent change in mean contraction period from baseline to treated), p-values from paired t-tests indicating statistical significance of the drug effect, and sample sizes (n = number of paired samples) noted. Note that these results are designed to be representative of what we can do with this pipeline. }
    
    \label{fig:pipeline} 
\end{figure}

Multiple computational frameworks have been developed to assess sarcomeric organization in cardiac myocytes. Some tools, such as ZlineDetection (Morris et al., 2020) and SarcOptiM (Pasqualin et al., 2016), are designed to extract averaged features from images of mature cardiac cells which exhibit well-organized sarcomeric structures. In contrast, other tools, such as SarcOmere Texture Analysis (Sutcliffe et al., 2018) were developed to address the challenge of analyzing reprogrammed myocytes derived from fibroblasts, which display less mature and more disorganized sarcomeric patterns similar to hiPSC-CMs. Subsequent approaches have advanced beyond averaged feature extraction to enable individual sarcomere segmentation and tracking. SarcTrack (Toepfer et al., 2019) introduced wavelet-based methods to segment and track sarcomeres in beating hiPSC-CMs, though often requiring substantial manual parameter tuning. More recently, SarcAsM (Härtter et al., 2025) leveraged deep learning for automated, multiscale segmentation and tracking of Z-bands, sarcomeres, and myofibrils in hiPSC-CMs. In 2021, our group introduced SarcGraph, an open-source pipeline with automatic detection and tracking of Z-discs and sarcomeres (Zhao et al., 2021; Mohammadzadeh and Lejeune, 2023; Mohammadzadeh and Lejeune, 2025).

Building on our previous work (Tsan et al., 2021), we adapted SarcGraph for the high-throughput analysis of high-frame-rate 2DMB videos. The original pipeline's reliance on sarcomere tracking proved unreliable for this dataset since 2DMBs exhibit much greater (and more physiologic) fractional shortening than iPSC-CMs cultured directly on glass (in our prior study, $5.0\pm1.1\%$ vs. $0.9\pm1.4\%$). The larger relative displacements in 2DMBs result in z-discs frequently moving in and out of the focal plane during contraction, as well as greater sarcomere displacement velocities, both of which caused inconsistent tracking with the original pipeline. Therefore, our modified workflow addresses this challenge by instead performing frame-by-frame sarcomere detection, automated tissue segmentation, and spatial partitioning to capture regional heterogeneity (Fig. 1a). From these measurements, an average sarcomere length time signal is computed and subsequently denoised using Gaussian process regression (GPR). A heuristic procedure then classifies the smoothed signal into relaxation, contraction, and recovery phases (Fig. 1b-i). To analyze spatial heterogeneity, the tissue mask is partitioned into predefined segments (e.g., longitudinal sections shown in the figure), and sarcomere analysis is performed within each subregion (Fig. 1b-ii). This pipeline yields an expanded suite of metrics quantifying sarcomere organization and contractile performance, supporting robust, paired comparisons across experimental conditions. For example, our analysis demonstrates that mavacamten induces a marked reduction in shortening amplitude that is highly consistent across paired measurements (Fig. 1c). Interestingly, both ET-1 and isoproterenol, which are known contractile agonists, decreased maximal sarcomere shortening amplitude but also caused a concomitant reduction in resting sarcomere length (Fig. 1c). These findings are consistent with greater sarcomere activation but would not be as readily discerned with whole tissue tracking. In contrast, mavacamten was associated with both reduced shortening amplitude and increased resting sarcomere length, consistent with reduced sarcomere activation (Fig. 1c). Furthermore, while all three treatments increased the frequency of contractions, isoproterenol and endothelin-1 produced greater reductions (26.5\% and 24.4\%, respectively) compared to mavacamten (10.4\%) (Fig. 1d).

For the analysis presented in this work (Fig. 1c-d), 65 paired baseline and treated samples were initially processed. From this set, 5 samples were excluded from the final analysis. These exclusions were necessary because the SarcGraph z-disc segmentation step failed specifically for the treated video of the pair. In these instances, instead of correctly identifying z-discs, the algorithm produced several thousand false positive detections, appearing to register signal noise as z-disc structures. We have not yet been able to pinpoint a clear relationship between this failure and any specific aspect of video quality that can be automatically identified before running SarcGraph on the samples. This remains an ongoing investigation that we aim to address in future work as we acquire more data samples.

Finally, we share \url{https://github.com/Sarc-Graph/sarcgraph-2dmb.git} under an open-source MIT license and make a dataset of 130 fluorescence example movies of 2DMB tissues publicly available via Harvard Dataverse to encourage collaborative efforts within the cardiac tissue engineering research community \url{https://doi.org/10.7910/DVN/GHMKWJ}. We anticipate that the availability of this image-based database, combined with our software pipeline, will not only facilitate the development of more robust computational tools for sarcomere-level analysis, but will also make adoption of the 2DMB framework significantly more accessible to others in the research community.

\section{Methods}
\label{sec:headings}

\subsection{Data}

Cardiomyocytes were differentiated from an $\alpha$-actinin-GFP reporter hiPSC line (Allen Institute) using previously described methods (Tsan et al., 2021). Following differentiation and metabolic purification, these hiPSC-CMs were seeded onto micropatterned 8 kPa substrates at day 22 following differentiation initiation to form cardiac muscle bundles (2DMBs). 2DMBs were matured for an additional 8 days with increasing calcium in the media by transitioning to a 75:25 DMEM:RPMI-B27 media, as previously described  (Tsan et al., 2021). Microscopy was performed with a Nikon Eclipse Ti-E inverted microscope with motorized stage. The 2DMBs were maintained in their standard culture media (75:25 DMEM:RPMI-B27) using a stagetop incubator to maintain temperature/humidity/C02. A Nikon Crest X-Light V2 spinning disk confocal light source was used to image sarcomeres (using the $\alpha$-actinin-GFP reporter) at 100 frames per second with a 40X long working distance water immersion Nikon objective. Three separate experiments were performed to assess the effects of known modulators of sarcomere function, mavacamten (300 nM), endothelin-1 (10 nM), and isoproterenol (50 nM). Images were collected at baseline and after 10 min treatment with each agent. Nikon Elements imaging analysis software was used to export files in the ND2 format. 

\subsection{Code}

The initial input for the analysis pipeline consists of an image sequence, provided as consecutive frames, capturing the contractile behaviour of a 2D cardiac muscle bundle at the z-disc level. Two primary processing scripts, \texttt{detection.py} and \texttt{segmentation.py}, are employed on this raw data. The \texttt{detection.py} script leverages SarcGraph (Mohammadzadeh and Lejeune, 2023), a previously published Python tool from our group, to detect individual z-discs and sarcomeres within each frame. For a detailed methodology on the SarcGraph algorithm, readers are referred to the original publications (Zhao et al., 2021; Mohammadzadeh and Lejeune, 2023; Mohammadzadeh and Lejeune, 2025). The \texttt{segmentation.py} script generates a tissue segmentation mask for each frame using a classical computer vision approach. This script first estimates background noise from a small region of the image to apply a noise-suppression threshold. The image is then smoothed using a Gaussian blur from the OpenCV library (\texttt{cv2.GaussianBlur}) and binarized using Otsu’s threshold, implemented in Scikit-image (\texttt{skimage.filters.threshold\_otsu}). A convex hull is then computed from the foreground pixels using SciPy (\texttt{scipy.spatial.ConvexHull}), and from these points, a minimum area bounding rectangle is fitted using OpenCV (\texttt{cv2.minAreaRect}) to define the tissue's boundary. This frame-by-frame segmentation is accelerated via Python's multiprocessing library. The algorithm may underestimate or overestimate the tissue area in cases of improper cropping where adjacent tissues are visible, or in the presence of artifacts like intense brightness variability where parts of the tissue are out of the focal plane. However, visual evaluation of representative samples confirmed this method was reliable for the dataset used in this work, and we anticipate that modifications for alternative datasets will either be unnecessary or straightforward to implement.

To enable regional analysis of contractile behavior, the \texttt{partition.py} script organizes the detected sarcomeres into distinct spatial segments. Using the per-frame sarcomere coordinates and their corresponding tissue bounding boxes, the script assigns two independent regional labels to each sarcomere. First, based on its position along the tissue's longer axis, a sarcomere is categorized as being in the 'left', 'center', or 'right' third of the tissue. Second, it is labeled as belonging to either the 'top' or 'bottom' half along the shorter axis. This labeling approach allows for subsequent analysis to be performed on either three longitudinal segments or a more granular six-region partition (e.g., 'left-top', 'center-bottom'). The final output is an annotated dataset where every sarcomere is tagged with its regional labels, facilitating a detailed comparison of contractility across different parts of the tissue.

Following regional partitioning, the raw sarcomere length data is processed to generate a clean time series for each selected tissue segment. First, for a given region (e.g., 'center' or 'whole tissue'), the average sarcomere length is computed across all sarcomeres for each frame, resulting in a time-series signal. To filter out noise inherent in imaging at this resolution, we employ a Gaussian Process Regression (GPR) model to denoise this signal. This approach is implemented using the GPyTorch library, which leverages GPU acceleration for computational efficiency (Gardner et al., 2018). GPR is a powerful method commonly used to model complex signals due to its flexible kernel functions. To capture the distinct features of cardiac contraction, our model uses a composite kernel. This kernel combines a Periodic component, which accounts for the cyclical nature of the beating signal, with RBF and Matern ($\nu=2.5$) components to accurately model the underlying complexities of the contraction waveform. The output of this step is the denoised time series of the average sarcomere length for the selected region.

After denoising, the \texttt{analyze.py} script implements a heuristic algorithm to parse the denoised sarcomere length time series into distinct physiological phases: a stable relaxed state, a contraction phase, and an expansion phase. The process begins by mean-centering the signal and calculating a smooth derivative of the signal using a Savitzky-Golay filter. The algorithm then proceeds as follows: First, it pinpoints the onset of contraction and the end of relaxation. This is achieved by analyzing the "positive blocks" where the sarcomere length is above average. Within these regions, a high-value zone is defined as the area where the signal's amplitude exceeds 80\% of the block peak. The algorithm then searches inward from the boundaries of this zone until the absolute value of the signal derivative falls below a user-defined \texttt{derivative\_threshold}. The two points identified by this search mark the start and the end of the relaxed state of the cell. Next, to identify the end of the contraction phase, the algorithm analyzes the "negative blocks". Within this interval, it locates the point of minimum sarcomere length. Together, these three identified time points fully delineate the contraction, expansion, and relaxed phases for subsequent analysis. Visual inspection of this heuristic's performance on our dataset confirmed that its results are robust, contingent upon successful prior denoising by the GPR model.

In the final stage of the pipeline, the \texttt{feature\_extraction.py} script uses the information generated in the preceding steps to compute a comprehensive set of quantitative metrics. These metrics are categorized into two groups: those describing the structural organization of the sarcomeres and the tissue, and those characterizing the functional properties of contraction. Details of these metrics are as follows:


\renewcommand{\arraystretch}{1.3}

\begin{longtable}{l p{0.6\textwidth}}
    
    \caption{Description of features extracted by the SarcGraph pipeline.} \label{tab:features} \\
    
    \toprule
    \textbf{Feature Name} & \textbf{Description} \\
    \midrule
    \endfirsthead
    
    \multicolumn{2}{l}{\textit{...Table \ref{tab:features} continued from previous page}} \\
    \toprule
    \textbf{Feature Name} & \textbf{Description} \\
    \midrule
    \endhead

    \bottomrule
    \multicolumn{2}{r}{\textit{Continued on next page...}} \\
    \endfoot

    \bottomrule
    \endlastfoot

    
    Number of Contractions & The total count of recorded full contraction events. \\
    Contraction Period & The average duration of a single beating cycle (in Seconds). \\
    Contraction Frequency & The average beating rate of the tissue (in Hz). \\
    Relaxed Sarcomere Length (Mean) & The mean sarcomere length ($\mu$m) during the relaxed state. \\
    Relaxed Sarcomere Length (Median) & The median sarcomere length ($\mu$m) during the relaxed state. \\
    Relaxed Sarcomere Length (25th Percentile) & The 25th percentile of sarcomere length ($\mu$m) during the relaxed state. \\
    Relaxed Sarcomere Length (75th Percentile) & The 75th percentile of sarcomere length ($\mu$m) during the relaxed state. \\
    Relaxed Sarcomere Length (Standard Deviation) & The standard deviation of sarcomere length ($\mu$m) during the relaxed state. \\
    Peak Sarcomere Length (Mean) & The mean sarcomere length ($\mu$m) at the peak of contraction. \\
    Peak Sarcomere Length (Median) & The median sarcomere length ($\mu$m) at the peak of contraction. \\
    Peak Sarcomere Length (25th Percentile) & The 25th percentile of sarcomere length ($\mu$m) at the peak of contraction. \\
    Peak Sarcomere Length (75th Percentile) & The 75th percentile of sarcomere length ($\mu$m) at the peak of contraction. \\
    Peak Sarcomere Length (Standard Deviation) & The standard deviation of sarcomere length ($\mu$m) at the peak of contraction. \\
    Shortening Amplitude & The average magnitude of sarcomere shortening, calculated as the difference between relaxed and peak-contracted average sarcomere lengths. \\
    Peak Shortening Velocity & The maximum speed ($\mu$m/s) of sarcomere length change during the contraction phase. \\
    Peak Lengthening Velocity & The maximum speed ($\mu$m/s) of sarcomere length change during the relaxation phase. \\
    Contraction Onset to Relaxation End Time & The total duration from the onset of contraction to the end of relaxation (in Seconds). \\
    Contraction Onset to Peak Contraction Time & The time taken to reach peak contraction from the onset (in Seconds). \\
    Contraction Onset to Half Contracted Time & The time taken to reach 50\% of maximal contraction from the onset (in Seconds). \\
    Half Contracted to Peak Contraction Time & The duration from half contraction to peak contraction (in Seconds). \\
    Peak Contraction to Relaxation End Time & The time taken to reach full relaxation from peak contraction (in Seconds). \\
    Peak Contraction to Half Relaxed Time & The time taken to reach 50\% relaxation from peak contraction (in Seconds). \\
    Half Relaxed to Full Relaxation Time & The duration from half relaxation to full relaxation (in Seconds). \\
    Tissue Length (Relaxed) & The tissue's overall length ($\mu$m) in the relaxed state. \\
    Tissue Length (Peak Contraction) & The tissue's overall length ($\mu$m) in the peak contracted state. \\
    Tissue Width (Relaxed) & The tissue's overall width ($\mu$m) in the relaxed state. \\
    Tissue Width (Peak Contraction) & The tissue's overall width ($\mu$m) in the peak contracted state. \\
    Tissue Area (Convex Hull, Relaxed) & The tissue's convex hull area ($\mu$m$^2$) in the relaxed state. \\
    Tissue Area (Convex Hull, Peak Contraction) & The tissue's convex hull area ($\mu$m$^2$) in the peak contracted state. \\
    Tissue Area (Bounding Box, Relaxed) & The tissue's bounding box area ($\mu$m$^2$) in the relaxed state. \\
    Tissue Area (Bounding Box, Peak Contraction) & The tissue's bounding box area ($\mu$m$^2$) in the peak contracted state. \\
    Total Number of Sarcomeres & The average number of individual sarcomeres detected per frame. \\
    Sarcomere Density (Convex Hull) & The average number of sarcomeres per unit of tissue area ($\mu$m$^2$) using the convex hull method. \\
    Sarcomere Density (Bounding Box) & The average number of sarcomeres per unit of tissue area ($\mu$m$^2$) using the bounding box method. \\
    Noise Level & A metric calculated as the standard deviation of the difference between the raw sarcomere length signal and the denoised signal. \\

\end{longtable}


\newpage
\section*{Reagents}

\begin{table}[h]
\centering
\begin{tabular}{l p{0.5\textwidth}}
    \toprule
    \textbf{REAGENT} & \textbf{AVAILABLE FROM} \\
    \midrule
    Sylgard-184 poly(dimethylsiloxane) (PDMS) & Dow Silicones Corporation \\
    Sylgard-527 PDMS & Dow Silicones Corporation \\
    Polyvinyl alcohol & Sigma-Aldrich \\
    Human fibronectin & Sigma-Aldrich \\
    6-cm and 6-well plate culture dishes & Various \\
    RPMI media & Gibco \\
    DMEM media & Gibco \\
    B27 supplement & Gibco \\
    Matrigel & Corning \\
    Endothelin-1 & Sigma-Aldrich \\
    Mavacamten & Caymen Chemical \\
    Isoproterenol & Sigma-Aldrich \\
    Dimethyl sulfoxide (DMSO) & Fisher \\
    \bottomrule
\end{tabular}
\end{table}

\begin{table}[h]
\centering
\begin{tabular}{l p{0.5\textwidth}}
    \toprule
    \textbf{CELL TYPE} & \textbf{DESCRIPTION, AVAILABLE FROM} \\
    \midrule
    $\alpha$-actinin-GFP hiPSC-derived cardiomyocytes & hiPSC line from Allen Institute (ID AICS-0075 cl.85) \\
    \bottomrule
\end{tabular}
\end{table}

\renewcommand{\arraystretch}{1.0}

\newpage
\section*{Acknowledgements}

This study was funded by National Science Foundation CELL-MET ERC EEC-1647837 (\url{https://www.nsf.gov/}) and the American Heart Association Career Development Award 856354 (\url{https://www.heart.org/}) awards to Emma Lejeune. The sponsors played no role in the study design, data collection and analysis, decision to publish, or preparation of the manuscript.

\newpage
\section*{References}

Bedada, F. B., Wheelwright, M., Metzger, J. M. 2016. Maturation status of sarcomere structure and function in human iPSC-derived cardiac myocytes. Biochimica et Biophysica Acta (BBA) - Molecular Cell Research, 1863(7 Pt B), 1829-1838. doi:10.1016/j.bbamcr.2015.11.005. PMID:26578113.

Burridge, P. W., Li, Y. F., Matsa, E., Wu, H., Ong, S. G., Sharma, A., et al., Wu, J. C. 2016. Human induced pluripotent stem cell-derived cardiomyocytes recapitulate the predilection of breast cancer patients to doxorubicin-induced cardiotoxicity. Nature medicine, 22(5), 547-556. doi:10.1038/nm.4087. PMID:27089025.

Ewoldt JK, DePalma SJ, Jewett ME, Karakan MÇ, Lin Y-M, Hashemian PM, et al., Chen CS. 2025. Induced pluripotent stem cell-derived cardiomyocyte in vitro models: benchmarking progress and ongoing challenges. Nature Methods. 22(1):24-40. doi:10.1038/s41592-024-02480-7. PMID:39516564.

Gardner, J. R., Pleiss, G., Bindel, D., Weinberger, K. Q., Wilson, A. G. 2018. GPyTorch: Blackbox Matrix-Matrix Gaussian Process Inference with GPU Acceleration. Advances in Neural Information Processing Systems, 31:7587-7597. doi:10.48550/arXiv.1809.11165.

Gerbin, K. A., Grancharova, T., Donovan-Maiye, R. M., Hendershott, M. C., Anderson, H. G., Brown, J. M., et al., Gunawardane, R. N. 2021. Cell states beyond transcriptomics: integrating structural organization and gene expression in hiPSC-derived cardiomyocytes. Cell Systems, 12(6), 670-687.e10. doi:10.1016/j.cels.2021.05.001. PMID:34043964.

Gintant, G., Burridge, P., Gepstein, L., Harding, S., Herron, T., Hong, C., et al., Wu, J. C. 2019. Use of human induced pluripotent stem cell–derived cardiomyocytes in preclinical cancer drug cardiotoxicity testing: a scientific statement from the American Heart Association. Circulation Research, 125(10), e75-e92. doi:10.1161/RES.0000000000000291. PMID:31533542.

Härtter, D., Hauke, L., Driehorst, T., Long, Y., Bao, G., Primeßnig, A., et al., Zimmermann, W.-H. 2025. SarcAsM: AI-based multiscale analysis of sarcomere organization and contractility in cardiomyocytes. bioRxiv. doi:10.1101/2025.04.29.650605.

Honda, Y., Li, J., Hino, A., Tsujimoto, S., et al., Lee, J.-K. 2021. High-throughput drug screening system based on human induced pluripotent stem cell-derived atrial myocytes ~ a novel platform to detect cardiac toxicity for atrial arrhythmias. Frontiers in Pharmacology, 12, 680618. doi:10.3389/fphar.2021.680618. PMID:34413773.

Huang, Y., Wang, T., López, M. E. U., Hirano, M., Hasan, A., Shin, S. R. 2020. Recent advancements of human iPSC derived cardiomyocytes in drug screening and tissue regeneration. Microphysiological Systems, 4, 2. doi:10.21037/mps-20-3. PMID:39430371.

Mohammadzadeh, S., Lejeune, E. 2023. SarcGraph: a Python package for analyzing the contractile behavior of pluripotent stem cell-derived cardiomyocytes. Journal of Open Source Software, 8(85), 5322. doi:10.21105/joss.05322.

Mohammadzadeh, S., Lejeune, E. 2025. Quantifying HiPSC-CM structural organization at scale with deep learning-enhanced SarcGraph. PLoS Computational Biology, 21(10), e1013436. doi:10.1371/journal.pcbi.1013436.

Mohr, E., Thum, T., Bär, C. 2022. Accelerating cardiovascular research: recent advances in translational 2D and 3D heart models. European Journal of Heart Failure, 24(10), 1778-1791. doi:10.1002/ejhf.2631. PMID:35867781.

Morris, T. A., Naik, J., Fibben, K. S., Kong, X., Kiyono, T., Yokomori, K., et al., Grosberg, A. 2020. Striated myocyte structural integrity: automated analysis of sarcomeric z-discs. PLoS Computational Biology, 16(3), e1007676. doi:10.1371/journal.pcbi.1007676. PMID:32130207.

Pasqualin, C., Gannier, F., Yu, A., Malécot, C. O., Bredeloux, P., Maupoil, V. 2016. SarcOptiM for ImageJ: high-frequency online sarcomere length computing on stimulated cardiomyocytes. American Journal of Physiology-Cell Physiology, 311(2), C277-C283. doi:10.1152/ajpcell.00094.2016. PMID:27335170.

Pioner, J. M., Guan, X., Klaiman, J. M., Racca, A. W., Pabon, L., Muskheli, V., et al., Murry, C. E. 2016. Isolation and mechanical measurements of myofibrils from human induced pluripotent stem cell-derived cardiomyocytes. Stem Cell Reports, 6(6), 885-896. doi:10.1016/j.stemcr.2016.04.006. PMID:27185283.

Sacchetto, C., Vitiello, L., de Windt, L. J., Rampazzo, A., Calore, M. 2020. Modeling cardiovascular diseases with hiPSC-derived cardiomyocytes in 2D and 3D cultures. International Journal of Molecular Sciences, 21(9), 3404. doi:10.3390/ijms21093404. PMID:32403456.

Sutcliffe, M. D., Tan, P. M., Fernandez-Perez, A., Nam, Y.-J., Munshi, N. V., Saucerman, J. J. 2018. High content analysis identifies unique morphological features of reprogrammed cardiomyocytes. Scientific Reports, 8(1), 1258. doi:10.1038/s41598-018-19539-z. PMID:29352247.

Toepfer, C. N., Sharma, A., Cicconet, M., Garfinkel, A. C., Mücke, M., Neyazi, M., et al., Seidman, C. E. 2019. SarcTrack. Circulation Research, 124(8), 1172-1183. doi:10.1161/CIRCRESAHA.118.314505. PMID:30700234.

Tsan, Y. C., DePalma, S. J., Zhao, Y. T., Capilnasiu, A., Wu, Y. W., Elder, B., et al., Helms, A. S. 2021. Physiologic biomechanics enhance reproducible contractile development in a stem cell derived cardiac muscle platform. Nature Communications, 12, 6167. doi:10.1038/s41467-021-26496-1. PMID:34697315.

Van den Berg, C. W., Okawa, S., Chuva de Sousa Lopes, S. M., van Iperen, L., Passier, R., Braam, S. R., et al. 2015. Transcriptome of human foetal heart compared with cardiomyocytes from pluripotent stem cells. Development, 142(18), 3231-3238. doi:10.1242/dev.123810. PMID:26209647.

Zhao, B., Zhang, K., Chen, C. S., Lejeune, E. 2021. Sarc-Graph: automated segmentation, tracking, and analysis of sarcomeres in hiPSC-derived cardiomyocytes. PLoS Computational Biology, 17(10), e1009443. doi:10.1371/journal.pcbi.1009443. PMID:34613960.

\end{document}